# Research-Based Course Materials and Assessments for Upper-Division Electrodynamics (E&M II)

Charles Baily, Michael Dubson and Steven J. Pollock

*Department of Physics, University of Colorado, Boulder, CO 80309-0390 USA*

**Abstract.** Favorable outcomes from ongoing research at the University of Colorado Boulder on student learning in junior-level *electrostatics* (E&M I) have led us to extend this work to upper-division *electrodynamics* (E&M II). We describe here our development of a set of research-based instructional materials designed to actively engage students during lecture (including clicker questions and other in-class activities); and an instrument for assessing whether our faculty-consensus learning goals are being met. We also discuss preliminary results from several recent implementations of our transformed curriculum, plans for the dissemination and further refinement of these materials, and offer some insights into student difficulties in advanced undergraduate electromagnetism.

## INTRODUCTION

There is substantial evidence from physics education research (PER) that introductory physics students learn and retain more when they are active participants in the classroom. [1, 2] Ongoing research at the University of Colorado (CU) and elsewhere has shown that *upper-division* students can likewise benefit from the use of in-class "clicker" questions and other student-centered activities. [1, 3-7] This work has demonstrated that active engagement can lead to increased learning in advanced physics courses (compared to standard lecture formats), and that junior-level *electrostatics* (E&M I) students are often still struggling with basic concepts. [8, 9] These important results have naturally motivated us to expand the context of this research at CU to advanced undergraduate *electrodynamics* (E&M II).

Faculty-consensus learning goals and research into common student difficulties have guided our development of a set of instructional materials and assessments for an active-learning electrodynamics curriculum. We outline here our transformation of an upper-division E&M II course for physics majors at CU, and provide details on the classroom activities we used, which were positively received by our students. We also compare preliminary assessment results from the recent implementation of this transformed curriculum at CU with another institution, and discuss future plans for research and refinement. All of these course materials (along with instructor guides, lists of learning goals, and observations of student difficulties) are freely available online, [10] and we encourage others to evaluate them, and to adapt them for their own use.

## TRANSFORMATION NARRATIVE

The *Science Education Initiative* model for course transformation [11] is an iterative process, where three key steps are used to inform all aspects of the project: establish explicit learning goals in collaboration with experienced faculty; apply education research to develop materials and teaching strategies to help students achieve those goals; use validated assessments to determine what students are (and aren't) learning.

We followed this model by first holding a two-day meeting in summer 2011 of 15 physics faculty members from a total of eight institutions (including CU), all having experience in PER and curriculum development, in order to brainstorm on student difficulties in advanced E&M, and to define our research goals. We found that the coverage of electrostatics was fairly standard across institutions, but topics from electrodynamics were often treated differently. At CU Boulder, electrodynamics is taught in the second half of a two-semester sequence, with classes of 30-40 students meeting for three 50-minute periods each week. Our usual text is Griffiths [12] (chapters 7-12), though instructors often add topics (e.g., *AC* circuits) or omit them according to preference. Physics majors at other institutions may instead cover most of advanced undergraduate E&M in a single semester, use a different textbook, and/or learn about wave optics and relativity in separate courses. To reach the greatest number of institutions, we decided to follow the presentation of topics in Griffiths, and to focus the assessment on core material likely to be covered in most electrodynamics courses.

This meeting was supplemented by individual interviews with six instructors who had recently taught E&M II at CU. We sought to understand how experienced physicist-teachers had approached this course in the past, what they felt were its essential elements, and their ideas on the particular challenges students face when the Maxwell equations become time-dependent. They also shared their homework problems, exams, lecture notes, and some clicker questions. Student interviews were held with five volunteers from the Spring 2011 (SP11) semester of E&M II, which confirmed many of the student difficulties these instructors had reported (e.g., trouble parsing the numerous vector quantities that appear in representations of electromagnetic plane waves). Our collaborations with non-PER faculty members at CU continued into Fall 2011 (FA11) with 3 one-hour meetings to establish explicit learning goals, and to vet assessment questions (see below).

We also arranged for ourselves to teach E&M II at CU in the FA11 [SJP/CB] and SP12 [MD/CB] terms. The instructors for both of these semesters typically used 3-5 clicker questions per class, interspersed throughout, comprising an estimated 20% of a 50-minute class period. Homework and exam questions rewarded reasoning and sense-making, along with traditional problem-solving skills. There were twice-weekly sessions outside of class where students could work together on homework (with occasional guidance from an instructor). Weekly online "preflight" questions [13] oriented students to upcoming topics, and their responses informed our class preparations. Homework and exams from every student were photocopied and archived for research purposes.

The FA11 course also occasionally used short, small-group activities to further engage students during class (e.g., working out an equation that might have otherwise been derived for them by an instructor at the board). These student-centered tasks (as well as other in-class discussions) were additionally facilitated by two undergraduate *Learning Assistants*, [14] who also met with us regularly outside of class to reflect on the course and discuss student difficulties.

To test the in-class tutorials under development, we recruited three FA11 students to participate in weekly group interviews throughout the semester. This interview format was chosen so we could observe the same students as they progressed through the course, simultaneously interacting with both the materials and each other. The interviewer mostly listened as students engaged with the tasks, occasionally asked clarifying questions, and provided guidance when needed (as would be typical of an actual tutorial setting). The activities were variously inspired by in-class observations, anticipated student difficulties, and *AJP* articles. [15] The topics were chosen to follow the lecture presentation, so as to capture students as they were first exposed to new material. We were thus able to gauge how students would interpret the wording and diagrams in the tutorials, the time required to complete them, and their overall usefulness for student learning.

These tutorials were modified for clarity and timing, and then used in the SP12 E&M II course at CU. About 40% of the lectures were partly or fully replaced with student-centered activities, each lasting from 10 to 50 minutes, and we typically had time to orient students to the upcoming tasks with a short lecture or series of concept tests. Audio recordings of single-group interactions were used to supplement our personal observations, which informed our subsequent revisions. The use of tutorials during the regular class period was fairly new to many of our students, and some initially needed additional encouragement to work with others (we found it helpful to remind them that scientific argumentation is a skill that is learned with practice). Overall, most aspects of the two transformed courses were extremely popular, with a large majority of students from both semesters rating them (in an end-of-term online survey) as either *useful* or *very useful* for their learning. [Fig. 1]

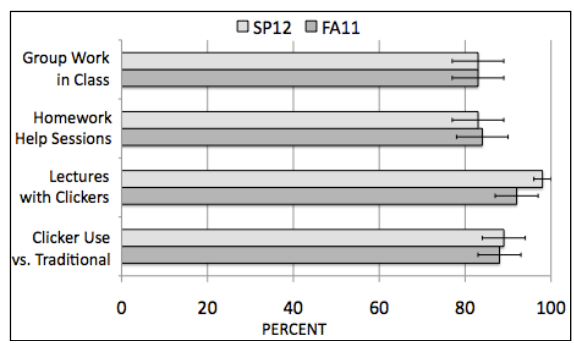

**FIGURE 1.** Percentage of SP12 (N=35) & FA11 (N=27) E&M II students at CU who rated aspects of the transformed courses as either *useful* or *very useful* for their learning. Error bars represent the standard error on the proportion.

## COURSE MATERIALS

We have compiled a suite of clicker questions, in-class activities, homework & exam problems, covering E&M topics from the second part of Griffiths; namely: the time-dependent Maxwell equations, conservation laws, EM waves in vacuum and media, potentials and gauge transformations, radiation, and special relativity (with additional material on *AC* circuits). This package of course materials [10] also contains implementation guides, an archive of several past E&M II courses at CU, and other supporting documents (e.g., explicit learning goals); along with source files (in PowerPoint and Word format) in order to facilitate their adaptation and implementation at other institutions.

## Learning Goals

This transformation process relied heavily on having explicit goals for student development; [10] the methods used for creating a list of broad and topic-specific learning goals have been described elsewhere in detail. [16] Our biggest issue was whether the broader student goals for E&M II (regarding their general development as physicists) would differ at all from those articulated for E&M I. One addition to this list concerns the increasingly mathematical nature of learning in advanced physics courses, particularly with electromagnetism as a classical field theory. The consensus of our working group was that students should understand the important role of formal mathematics in learning and applying physics; more specifically, know how the assumptions made when deriving an equation define its range of validity.

## Clicker Questions

Clicker questions can be effectively incorporated into upper-division classrooms in a variety of ways, [3] though they're most often associated with gauging *conceptual understanding* of newly presented material (i.e., questions that don't require mathematical calculation). [Fig. 2] They might also be used, for example, to underscore an essential point in the middle of a long derivation, to have students apply results to a novel situation, or to make direct connections between mathematical equations and the physical situations they describe. Further details are given in the package of course materials, where the clicker questions are annotated with prior student responses, instructor notes, and comments on the in-class discussions they inspired. [10]

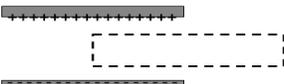

**FIGURE 2.** A *conceptual* clicker question that inspired important in-class discussions. 30% of FA11 E&M II students at CU were initially incorrect, either forgetting the existence of fringe fields, or not seeing how these contributed to the line integral. Even knowing that the EMF should be zero here, many students could not argue for this using Maxwell's equations.

## Assessment & Preliminary Results

The *Colorado UppeR-division ElectrodyNamics Test* (CURrENT) is an assessment of fundamental skills and understanding in core topics from advanced undergraduate electrodynamics. The basic (though **not** introductory-level) nature of these six multi-part questions reflects our premise that a more sophisticated understanding of advanced E&M is unlikely for students who haven't yet mastered essential concepts. Its open-response format follows from an expectation that more advanced students should be able to generate their own answers, and to justify their correctness. The focus is *conceptual*, though some mathematical manipulations are required (per our learning goals); in particular, Q4 asks students to transform a curl equation to its integral form via Stokes' theorem. More typical of the assessment would be Q3, which asks whether the E-field just outside of a current-carrying wire is *zero* or *non-zero*; and likewise regarding the divergence of the steady-state current density inside the wire. [See Fig. 3]

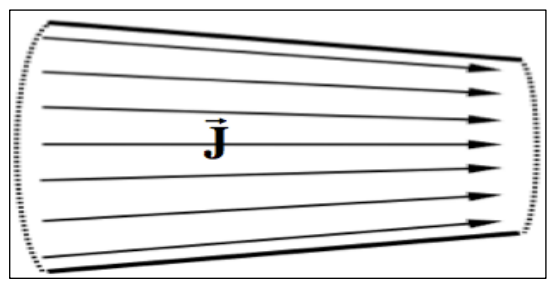

**FIGURE 3.** Diagram from Q3 of the CURrENT, showing a radially decreasing wire that carries a steady current density **J**. The parallel components of the E-field are continuous across any boundary, making the field non-zero just outside the surface of the wire. Div.**J** = 0 inside the wire by charge conservation (or, the continuity of the field lines).

The CURrENT was given in SP12 at CU (N=24), and also at a small, selective engineering college ("X"; N=11). Instructor X was present at our summer meeting, had access to the FA11 CU course materials, but mostly used locally-created clicker questions and preflights. Students in both courses were encouraged to not study for this ungraded assessment, and instead use it to judge their understanding before preparing for the upcoming final exam. Responses from both institutions were scored using a consistent rubric, and the average CU SP12 scores were significantly higher for Q3 and Q4 ($p \leq 0.001$), and for the total score ($p < 0.05$). [Fig. 4] Scores were not significantly different for the other questions, though X-students performed better than CU on Q2, regarding the fields produced by a time-varying solenoidal current.

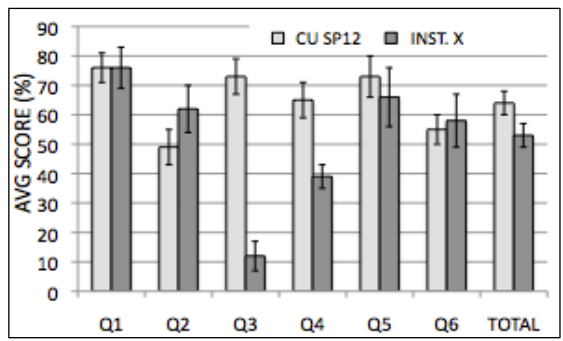

**FIGURE 4.** Average CURrENT scores (by question and total) for CU SP12 (N=24) and Institution X (N=11). Significant differences are seen in Q3 & Q4 ($p \leq 0.001$) and the total score ($p < 0.05$). Error bars represent the standard error on the mean.

The relatively high CU SP12 results for Q4 can be attributed (at least in part) to our emphasis on the equivalency of the differential and integral forms of Maxwell's equations, and on the coordination of lines, surfaces and volumes when applying the integral forms. 13/24 of our students were able to correctly and completely explain each of their three steps in this short derivation; only 3/11 X-students reached the final result, though none of them offered complete and correct reasoning for each and every step. Instructor X reported that he hadn't reviewed the Divergence and Stokes' theorems since the first semester.

Q3 can be answered without any calculations, but requires an understanding of the microscopic version of Ohm's law (**J** = σ**E**), boundary conditions on electric fields, and conservation of charge. These topics are directly addressed in many of the CU clicker questions (as well as several of the tutorials), and most SP12 CU students answered both parts of this question correctly. However, 10/11 X-students *incorrectly* thought $\nabla \cdot \mathbf{J}$ would be non-zero inside the wire (the remaining student left this question blank), because the magnitude of **J** is increasing to the right. At least half of them were distracted by the appearance of "converging" field lines, which they took to represent a non-zero divergence. Only 1/11 X-students could correctly explain why the electric field just outside the current-carrying wire should be non-zero (the parallel components of the E-field are continuous across any boundary, as required by Faraday's law). Student reasoning was varied here, though several incorrectly applied Gauss' law to argue that the electric field in the region just outside the wire is zero because the charge density there is zero. This is consistent with our observation that E&M II students may still sometimes think that a vanishing divergence (or line-integral) of a field implies that the field itself is zero.

## FUTURE STEPS

The richness of student responses to both the assessment and tutorial questions indicate the promise of these materials as tools for research into student learning in advanced E&M, and we will continue to collect data to evaluate their use and reliability. Modifications to the in-class activities and assessments will be made following their implementation at CU and elsewhere in the FA12 semester. The latest version of the CURrENT is scheduled to be given at another institution (different from "X"), where the course will be taught by a PER instructor. We encourage instructors elsewhere to adopt these materials and/or give our assessments in their own courses, and to share with us their results and reflections. This would allow us to make further comparisons of different teaching approaches and student populations, and to assess the effectiveness of this active-learning curriculum relative to traditional modes of instruction.

## ACKNOWLEDGMENTS

We gratefully acknowledge the contributions of the many students and faculty who participated in this project, particularly: B. Ambrose, P. Beale, E. Kinney, F. Kontur, T. Munsat, S. Ragole, D. Rehn, A. M. Rey, C. Rogers, E. Zimmerman and the members of the PER@C group. This work supported by the University of Colorado and NSF-CCLI grant #1023028.